\documentclass[12pt]{article}
\usepackage{latexsym}
\usepackage{epsfig}
\usepackage{graphicx}
\usepackage{epstopdf}

\usepackage{epsfig,amssymb,amsmath,amscd}

\newcommand{\bq}{\begin{equation}}
\newcommand{\eq}{\end{equation}}
\newcommand{\bqa}{\begin{eqnarray}}
\newcommand{\eqa}{\end{eqnarray}}
\newcommand{\ben}{\begin{enumerate}}
\newcommand{\een}{\end{enumerate}}
\newcommand{\bc}{\begin{center}}
\newcommand{\ec}{\end{center}}
\newcommand{\bqb}{\begin{eqnarray*}}
\newcommand{\eqb}{\end{eqnarray*}}

\def\pr#1#2#3{Phys. Rev. ${\bf{#1}}$, #2 (#3)}
\def\prl#1#2#3{Phys. Rev. Lett. ${\bf{#1}}$, #2 (#3)}

\def\epj#1#2#3{Eur. Phys. J. ${\bf{#1}}$, #2 (#3)}

\def\jmp#1#2#3{J. Mod. Phys. ${\bf{#1}}$, #2 (#3)}

\begin{document}
\pagenumbering{arabic}
\thispagestyle{empty}
\def\thefootnote{\fnsymbol{footnote}}
\setcounter{footnote}{1}

\begin{flushright}
Jan.17, 2017\\
arXiv: \\
 \end{flushright}

\begin{center}
{\Large {\bf Higgs boson compositeness:\\
 from $e^+e^-\to  ZH$ to $e^+e^-\to Z_L,W_L +anything $}}.\\
 \vspace{1cm}
{\large F.M. Renard}\\
\vspace{0.2cm}
Laboratoire Univers et Particules de Montpellier,
UMR 5299\\
Universit\'{e} Montpellier II, Place Eug\`{e}ne Bataillon CC072\\
 F-34095 Montpellier Cedex 5, France.\\
\end{center}

\vspace*{1.cm}
\begin{center}
{\bf Abstract}
\end{center}

We assume that the Higgs doublet has a composite structure,
respecting the main standard model properties, and therefore called 
Composite Standard Model (CSM), but leading (through Goldstone equivalence) 
to $Z_L$ and $W_L$ form factors.
We illustrate how such a form factor affecting the $ZZ_LH$
coupling will be directly observable in $e^+e^-\to  ZH$.
We then show the
spectacular consequences which would appear in the inclusive processes
$e^+e^-\to Z_L+anything $.
Such a form factor could also affect the $\gamma W^+_LW^-_L$ and
$Z W^+_LW^-_L$ vertices and we show what effects this would generate in
$e^+e^-\to  W^+W^-$ (especially in $e^+e^-_R\to  W^+_LW^-_L$)
and in $e^+e^-\to W_L+anything $. 
We finally mention the $\gamma\gamma\to  W^+W^-$ process and several other 
processes in hadronic collisions which could be similarly affected.

\vspace{0.5cm}

\def\thefootnote{\arabic{footnote}}
\setcounter{footnote}{0}
\clearpage

\section{INTRODUCTION}

Higgs boson compositeness is an appealing possibility for understanding
the peculiar features of the standard model \cite{Hcomp}. Compositeness
could even concern other sectors; substructures had been considered
since a long time, see for ex.  \cite{comp}, \cite{Portal}, \cite{Htop},
\cite{BSMth}.
Several processes have been studied in order to test Higgs boson compositeness
or the fact that the Higgs boson could be a portal to new sectors (possibly
involving unvisible states), see ref.\cite{Hincl,HHH,Htt}.\\
In this paper we assume that Higgs boson compositeness
reproduces (possibly in an effective way) the general features of the SM,
in particular those resulting from gauge invariance and Goldstone equivalence.
At low energy no anomalous coupling would be generated.
As the energy increases we keep Goldstone equivalence which immediately ensures
(at scale $m_Z$) a good behaviour of the $W_L$, $Z_L$ amplitudes owing to
the typical SM cancellations.
But this means that $W_L$, $Z_L$ amplitudes will be equivalent
to the composite $G^{\pm},G^0$ ones and should reflect their
compositeness properties, possibly the presence of a form factor
(similarly to the hadronic case) with a new physics scale $M$.
Such a form factor should be close to 1 at low $q^2$ but, after showing some structures
around the new physics scale ($q^2\simeq M^2$), it may decrease at very high  $q^2$.
We will call models leading to such properties as Composite Standard Models (CSM).\\
Concerning the presence of a form factor in the $ZZ_LH\simeq ZG^0H$ coupling, the $e^+e^-\to ZH$ process, studied theoretically and experimentally
since a long time (see reviews and references in \cite{ZH1,ZH2}) should be particularly
interesting as it is largely dominated by $e^+e^-\to Z \to  Z_LH$ at high energy,
see \cite{SRSZH}.
This form factor could then be directly measured by the
$e^+e^-\to ZH$ cross section.
We illustrate this possibility with a test form factor controlled
by a new physics scale M.\\
This process will then furnish the basic source of an input for predicting CSM
effects in other processes. As an example we treat the case of the
inclusive process $e^+e^-\to Z_L+anything$.\\
We then look at the charged $W_L$ case. We start with the simplest
example given by $e^+e^-\to W^+W^-$ .
A high energy description has been given in SM (and in MSSM) in ref.\cite{SRSWW}.
The Helicity Conservation rule (HC), ref.\cite{hc}, predicts the presence of four 
leading amplitudes, two for $W^+_TW^-_T$ and two for $W^+_LW^-_L$.
The transverse amplitudes are larger than the longitudinal ones
but have different angular dependences.
So we study the importance of the $W^+_LW^-_L\simeq G^+G^-$ contribution versus energy and angle,
the effects of the presence of $\gamma G^+G^-$ and  $Z G^+G^-$
form factors and their  observability through the $e^+e^-\to W^+W^-$ cross section
with polarized or unpolarized $e^{\pm}$ beams.
We then also treat the case of the
inclusive process $e^+e^-\to W_L+anything$.\\
We finally mention other processes in which similar sudies may be done.
We say a few words about $\gamma\gamma \to W^+W^-$ .
More complex processes with $ZZ,WW,ZW$ production in hadronic collisions
could also be studied.

Section 2 is devoted to the description of the $e^+e^-\to ZH$ process and the possible
determination of the form factor.
The $e^+e^-\to Z_L+anything$ process is then studied in Section 3.
The charged cases $e^+e^-\to W^+W^-$ and
$e^+e^-\to W_L+anything$ are treated in Sections 4,5 and the
$\gamma\gamma \to W^+W^-$ process in Section 6..
The summary and the possibility of  other applications are discussed in Section 7.\\

\section{THE $e^+e^-\to ZH$ PROCESS AS THE BASIC\\
 SIGNAL OF HIGGS COMPOSITENESS}
The high energy properties of this process have been given in ref.\cite{SRSZH}.
At Born level the helicity amplitudes
consist in four $Z$-Transverse amplitudes ($\lambda=\pm{1\over2}$, $\tau=\pm1$)
\bqa
F^{Born}_{\lambda,\tau}&=& -~{e^2f_{ZZH}\sqrt{s}\over \sqrt{2}(s-m^2_Z)}
[g^Z_{eL}(\tau\cos\theta-1)\delta_{\lambda,-}-g^Z_{eR}(\tau\cos\theta+1)\delta_{\lambda,+}]
\eqa
and two $Z$-longitudinal amplitudes $\lambda=\pm{1\over2}$, $\tau=0$)
\bqa
F^{Born}_{\lambda,0}&=&~{e^2f_{ZZH}E_Z\sqrt{s}\sin\theta\over m_Z(s-m^2_Z)}
[g^Z_{eL}\delta_{\lambda,-}-g^Z_{eR}\delta_{\lambda,+}]
\eqa
At high energies the (HCns) rule \cite{hc} requires
\bq
\lambda +\lambda'=\tau ~~. \label{heli-cons}
\eq
and indeed the four transverse amplitudes which violate this rule (and are called HV) are
suppressed like $m_Z/\sqrt{s}$.
On the opposite the two longitudinal ones, which satisfy this rule (and are called HC)
tend to constants
\bqa
F^{Born}_{\lambda,0}&\to&~{e^2f_{ZZH}\sin\theta\over2m_Z}
[g^Z_{eL}\delta_{\lambda,-}-g^Z_{eR}\delta_{\lambda,+}]
\eqa
These HC amplitudes agree with the direct computation of the Goldstone process $e^-e^+\to G^0H$,
and the relation $F^{Born}(Z_0H)=iF^{Born}(G^0H)$ when neglecting ${m_Z^2\over s}$ terms.\\

In ref.\cite{SRSZH} the one loop contributions had also been computed and the explicit
high energy expressions in the so-called SIM approximation have been given in its Appendix A.\\

As explained in the Introduction, we will assume that Higgs compositeness generates
an s-dependent form factor $F_H(s)$ for the $ZG^0H$ coupling. So we can write, at Born
and at one loop SIM level

\bqa
F^{Comp}_{\lambda,0}\simeq F^{SIM}_{\lambda,0}F_H(s)
\eqa

The cross section given by

\bq
{d\sigma\over d\cos\theta}={\beta_Z\over 128\pi s}
\sum_{\lambda \tau}|F_{\lambda \tau}(\theta)|^2 ~~, \label{dsigma-unpol}
\eq
should allow to measure the form factor $F_H(s)$.
A first step could consist in neglecting the small HV amplitudes in which case
one gets simply
\bq
{d\sigma^{Comp}\over d\cos\theta}\simeq{d\sigma^{SIM}\over d\cos\theta}|F_H(s)|^2
\eq
such that $|F_H(s)|^2$ can immediately be obtained from the ratio
of the measured cross section over the predicted SM one.
A more precise result can then be obtained by taking also into account
the contribution of the small HV amplitudes.\\
In order to illustrate these properties we take a test form factor
\bq
F_H(s)={(m_Z+m_H)^2+M^2\over s+M^2}
 \label{FF}\eq
where $M$ is a new physics scale taken as $0.5$ TeV. Larger values of this
scale would require higher energies to see similar effects. \\
In Figure 1 (upper panel) we can first see how much the $Z_L$ production dominates
the unpolarized cross section and then how its reduction due to the form factor
affects it in a similar way.
In the lower panel, for $\sqrt{s}=4$ TeV, we show how the angular distribution (dominated by the
$\sin^2\theta$ shape of $Z_L$ production) is affected by the form factor.\\
Such a form factor which would be determined by this measurement
could then constitute the basic input for predicting corresponding
effects of compositeness in other processes.\\
Another particularly interesting process would be $\gamma\gamma \to ZH$
in which again the leading (HC) amplitude only involves $Z_LH$ and is therefore 
directly sensitive to the considered compositeness effects. This amplitude starts 
however at one loop (see for example \cite{ggVH}) and a careful analysis is required.\\
In the next Section we discuss the inclusive process $e^+e^-\to Z_L+anything$.

\section{THE PROCESS $e^+e^-\to Z_L+anything $}

At high energy and at leading order,  with $Z_L\simeq G^0$,
the SM contributions (apart from the above 2-body $G^0H$ process)
consist in the 3-body processes $G^0t\bar t$, $G^0HH$, $G^0ZZ$,
$G^0WW$ and $G^0ZH$, $G^0\gamma H$ channels.
The first 4 channels proceed through $e^+e^-$ annihilation into
photon or $Z$ followed by the 3-body production.
The last 2 ones proceed respectively through $e^+e^-\to ZZ$
and $e^+e^-\to \gamma Z$, followed by $Z\to G^0H$.\\
We have computed the corresponding inclusive cross section

\bq
{d\sigma\over dxdcos\theta}
\eq
\noindent
where $x={2p\over\sqrt{s}}$ is the reduced $Z_L$ momentum,
for fixed $\theta$ angle with respect to the $e^-$ direction; $s=q^2=(p_{e^+}+p_{e^-})^2$.\\

We then affect the vertices involving $H$ or $G^0$ by the above $F_H(s)$ form factor 
adapting the variable $s$ to the corresponding virtual momentum squared.
For simplicity we take the same form factor in all these cases. This is
arbitrary but our aim is not to make a precise prediction but to see what
type of modification could appear in the inclusive $x$ distribution.\\
In Figure 2 one can compare the resulting "SM" and "SMFF" curves for $\sqrt{s}=$ 4 TeV
with 2 choices of new physics scale, M=0.5 and 2 TeV; the
form factor effect indeed consists in a strong reduction of the cross section.\\

For comparison we have also drawn in Fig.2 the $Z_L\simeq G^0$ inclusive
distribution with the effect of new channels and with a crude parton-like
effect typical of compositeness.
We have first considered the production of pairs of new particles ($G^{'0}+H'$) one of these
particles emitting the final $G^0$ (for example $G^{'0}\to G^0+H$). We illustrate the effects
of 3 such pairs with common masses $M_{1,2,3}=0.5,1,1.5 TeV$.
The corresponding shapes result from the phase space and from the internal propagator
effects.
These are pure "kinematical shapes" and do not correspond to precise models
which may contain further effects due to precise intermediate states or resonances.
As already mentioned our aim is just to illustrate the sensitivity of the inclusive distribution
to the presence of new contributions.\\
We then show the shape of a parton-like distribution generated like in
the hadronic case by the following structure:
\bq
{d\sigma\over dxdcos\theta}=\Sigma_i{d\sigma_i\over dcos\theta} D_i(x)
\eq

We make an arbitrary illustration choosing as basic production cross section
${d\sigma_i\over dcos\theta}$ the standard
$e^+e^-\to ZH$ process then followed by
a normalized fragmentation function
\bq
\int^1_{1-{M^2\over s}} xD(x)dx=1
\eq
with
\bq
D(x)={6\over(1-{M^2\over s})^3}(1-x-{M^2\over s})
\eq
which favours the low x domain ($x<1-{M^2\over s}$
corresponding to a set of new states with a mass larger than $M$).\\
Let us add that in SM the fraction of $e^+e^-\to Z_L+anything $ production 
within the unpolarized $e^+e^-\to Z+anything $ case depends on the
energy and on the kinematical detection cuts, but should only be of the order
of 10 percent. A small new physics signal only located in $Z_L$ would then 
not be immediately observable in the unpolarized case and would
necessarily require a final Z polarization analysis.\\

\section{CONSEQUENCES FOR $e^+e^-\to W^+W^-$}

We now look at the charged $W_L$ case.
We will start with the simplest process, $e^+e^-\to W^+W^-$, observable at high energy
at a future linear collider, see ref.\cite{eeWWLC} and we will then look at the
inclusive process $e^+e^-\to W_L+anything$.\\

At Born level, as described in ref.\cite{SRSWW}, $e^+e^-\to W^+W^-$ is due to
neutrino exchange in the t-channel and photon+Z exchange in the s-channel, which
lead at high energy to two HC Transverse-Transverse (TT)  amplitudes ($\mu=-\mu'=\pm1$)
\bq
F^{\rm Born}_{-{1\over2}\mu-\mu}\to - {e^2 \sin\theta (\mu-\cos\theta)
\over 4s^2_W (\cos\theta-1)}~~. \label{Born-asym-TT-HC}
\eq
and two HC Longitudinal-Longitudinal (LL) amplitudes  ($\mu=0,\mu'=0$)
\bqa
F^{\rm Born}_{-{1\over2}00} & \to & - {e^2\over8s^2_Wc^2_W} \sin\theta ~~,
\nonumber \\
F^{\rm Born}_{+{1\over2}00} & \to & {e^2\over4c^2_W} \sin\theta ~~, \label{Born-asym-LL-HC}
\eqa
which both tend to constant values. On the opposite the other (TT,LL or TL) amplitudes
which are helicity violating (HV) vanish at high energy.\\
The one loop corrections to the HC amplitudes are also given in the SIM
approximation in ref.\cite{SRSWW}.
It had been checked that the LL amplitudes agree with equivalence to
$e^+e^-\to \gamma, Z \to G^+G^-$ amplitudes.\\
Note the dependence on the $e^+e^-$ polarization (with always $\lambda_{e^-}=-\lambda_{e^+}$).
The left case ($\lambda_{e^-}=-{1\over2}$) contributes to both TT amplitudes and to one LL amplitude,
whereas the right one ($\lambda_{e^-}=+{1\over2}$) only contributes to the other LL amplitude.\\
We will now assume that the $\gamma G^+G^-$ and  $Z G^+G^-$ vertices are
affected by the same type of form factor $F_H(s)$ as the above $ZG^0H$ vertex.
This will concern the two LL amplitudes. The two TT amplitudes should not be
affected. As these TT amplitudes are more important than the LL ones,
and depend differently on the angle, in the unpolarized case one does not get a simple factorization
of the form factor effect.\\
Only in the right-handed $e^+e^-$ polarization case involving only one LL amplitude
one gets this factorization at high energy.\\
Hence we have computed both the unpolarized differential cross section
\bq
{d\sigma\over d\cos\theta}={\beta_W\over 128\pi s}
\Sigma_{\lambda \mu \mu'}|F_{\lambda \mu \mu'}(\theta)|^2 ~~, \label{dsigma-unpol}
\eq
as well as the polarized differential cross section using  right-handedly polarized
 electron beams $e^-_R$,
\bq
{d\sigma^R\over d\cos\theta}={\beta_W\over 64\pi s}
\Sigma_{\mu \mu'}|F_{+{1\over2},\mu \mu'}(\theta)|^2 ~~, \label{dsigma-pol}
\eq
with and without form factor effects.\\
In the illustrations we have chosen the same test form factor as in the above $ZH$ case, eq.(\ref{FF}).\\

Figure 3 (upper panel) shows the energy dependence of the unpolarized cross section and of the
right polarization case. The lower panel shows their angular dependence at 4 TeV.
Indeed one sees that the form factor effect leads to only few percent effect
on the unpolarized cross section but generates a very large reduction of the
right-handed polarized one.
The angular dependencies confirm these effects.
So $e^+e^-_R\to W^+W^-$, dominated by $e^+e^-_R\to W^+_LW^-_L$, should be a favoured
place for observing this effect.\\

\section{THE PROCESS $e^+e^-\to W_L+anything $}

We now make a study of the inclusive  $e^+e^-\to W_L+anything $ process similarly to
the above $e^+e^-\to Z_L+anything $ case.\\
Assuming the $W^{\pm}_L\simeq G^{\pm}$ equivalence we discuss
the contributions to the inclusive process $e^+e^-\to G^- +anything $.
The basic SM terms (apart from the 2-body $e^+e^-\to G^- +W^+$)
corrrespond to  $G^-t\bar b$, $G^-HW^+$, $G^-ZW^+$ and $G^-\gamma W^+$ final states.
The $G^-t\bar b$ case proceeds via $e^+e^-$ annihilation into
photon or $Z$ followed by the 3-body production. The other 3 cases proceed, in addition,
via $e^+e^-\to W^+W^-$ (via t- and s- channel exchanges) followed by $W^-\to G^-H, G^-W,G^-\gamma$.\\
We compute the resulting inclusive distributions, first with SM couplings
and secondly with $G^-$ and $H$ couplings affected by the above $F_H(s)$ form factor.
Results can be seen in Figure 4 showing the important suppression of the SMFF cases
(as in Figure 2), especially at large $x$.\\
We have then compared, like in the $Z_L$ case, the effects of new particle production
and of a parton like distribution.
The different shapes are globally similar to the ones observed in $e^+e^-\to Z_L+anything $
and confirm the power of such inclusive distributions for looking at possible
compositeness signals.\\
We should also mention that in SM the $e^+e^-\to W_L+anything $ production, depends strongly
on the kinematical conditions, and may only be of the order  
of 10 percent of the unpolarized $e^+e^-\to W+anything $ case, so that a clear observation
of new effects may require a final W polarization analysis.\\

\section{AN OTHER EXAMPLE WITH $\gamma\gamma \to W^+W^-$ }

This process should be observable at a future $\gamma\gamma$ collider, for a recent review see \cite{gammagamma}.\\  
At Born level it is described by 3 diagrams: two ones for $W$ exchange in the $t$, $u$
channels and one corresponding just to the four body $\gamma\gamma W^+W^-$ coupling. At high energy, among the 16 helicity amplitudes, the leading ones are the four TT ones, ($----, ++++, -+-+, +-+-$), and the two LL ones, ($-+00, +-00$). One can check that at high energy, in the case of the LL amplitudes, very precise cancellations occur among the contributions of the
three diagrams in order to avoid increasing behaviours violating unitarity. The final result agrees with the computation of the $\gamma\gamma \to G^+G^-$ process
with similar three diagrams. This agreement does not occur separately for each diagram, but only after their addition as expected from the gauge invariance of the total amplitude.\\
One may now consider the effect of $G^{\pm}$ compositeness, i.e. a reduction at
high energy of these LL contributions due to form factors in the Goldstone couplings.\\
However, numerically, as it can be seen in Figure 5, it appears that, in SM, the LL amplitudes are about 4 times smaller than the TT ones and finally the LL contribution to the cross section is only of about 5 percent and probably unobservable.\\
So in order to observe this strong modification of the LL contribution due to compositeness effects (also shown in Figure 5) a final $W$ polarization analysis would be required
in order to only deal with $W^+_LW^-_L$ production.\\

\section{CONCLUSIONS AND FURTHER DEVELOPMENTS}

In this paper we have assumed that the Higgs doublet has a
compositeness origin which reproduces in an effective way
the main low energy SM features but may generate form factors
which affect the Higgs couplings at high energy.
Assuming that the Goldstone equivalence is also maintained
we expect that the $Z_L\simeq G^0$ and $W^{\pm}_L\simeq G^{\pm}$ couplings
will be affected in the same way (the CSM hypothesis).\\
We proposed to test these effects in simple $H$, $Z_L$ and $W_L$
production processes.\\
We have shown that the $e^+e^-\to  ZH$ process, largely dominated by
$e^+e^-\to Z \to  Z_LH$ is directly sensitive to the $ZZ_LH$ coupling
and to the presence of a form factor. Illustrations with an arbitrary
choice of form factor have been given. This process could then furnish
the basic input for further predictions.
We have illustrated the case of $e^+e^-\to Z_L+anything $ which can show
spectacular effects.\\
In a similar fashion we have then treated the charged case, i.e. the
compositeness effects of $G^{\pm}\simeq W^{\pm}_L$.
The simplest case is  $W^+W^-$ production in $e^+e^-$ collision and we have also
considered the inclusive case $e^+e^-\to W_L+anything $. We have finally looked at
the $\gamma\gamma\to W^+W^-$ process. In order to get large signals one needs to
isolate the polarized $W_L$ components.\\
We emphasize the fact that, in SM,
$e^+e^-\to W^+_LW^-_L$  is dominant at high energy by using right-handedly 
polarized $e^-$ beams and could immediately give a signal of $W^{\pm}_L$ compositeness.\\
Other ways of $Z,W,H$ production could also be considered
at hadronic colliders for example $ZH,WH,ZW,WW$ through gluon-gluon or $q\bar q$
collisions. Again a particular case is $gg \to ZH$ dominated by $Z_LH$ (\cite{ggVH})
but occuring at one loop.
Properties of $Z_L$ and $W_L$ could also be studied through
$Z_LZ_L,W_LW_L,Z_LW_L$ scattering but this deserves specific involved
studies, see for example reviews in \cite{VBS,VV1,VV2}.\\

\newpage

\begin{figure}[p]
\[
\epsfig{file=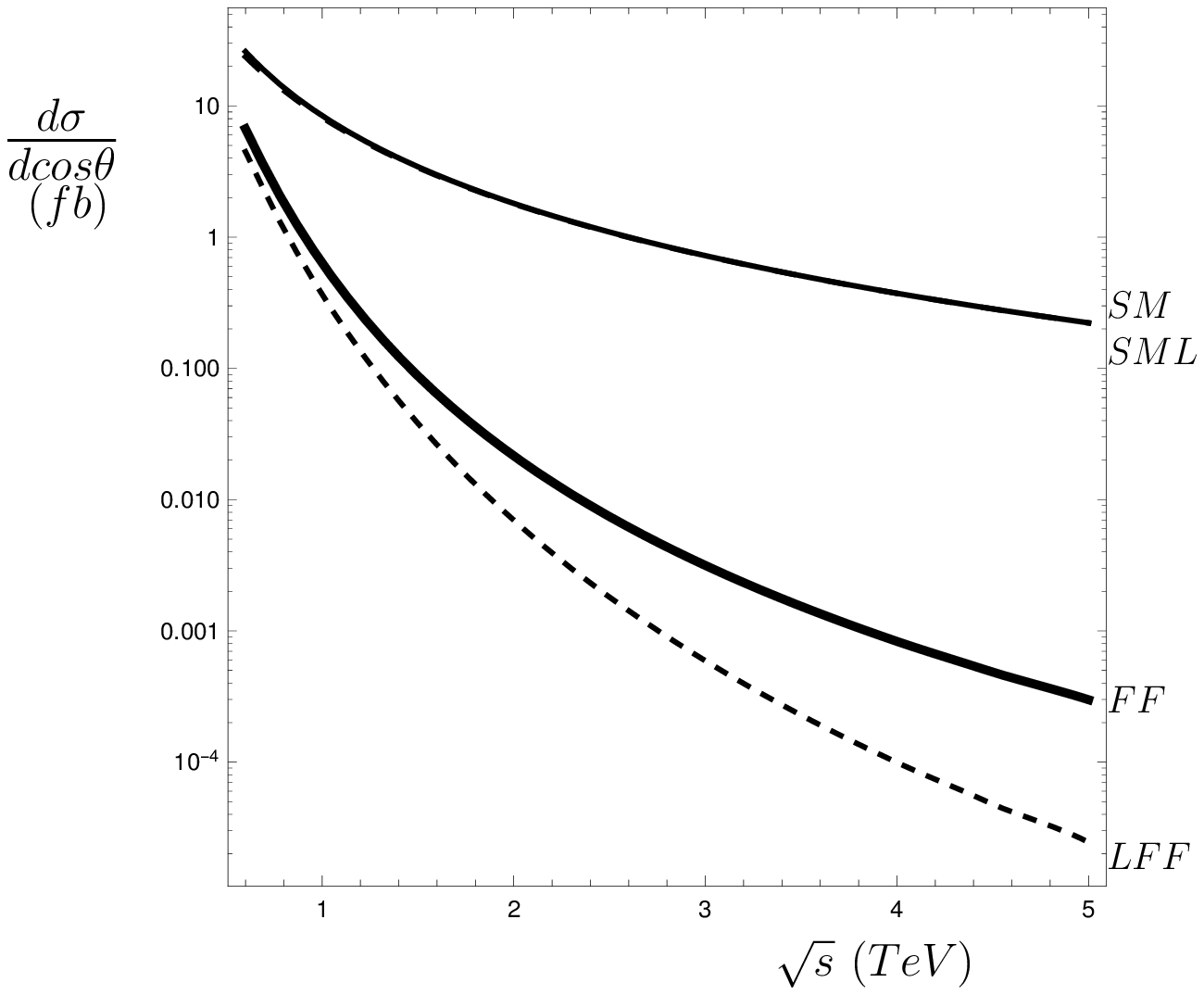, height=8.cm}
\]\\
\[
\epsfig{file=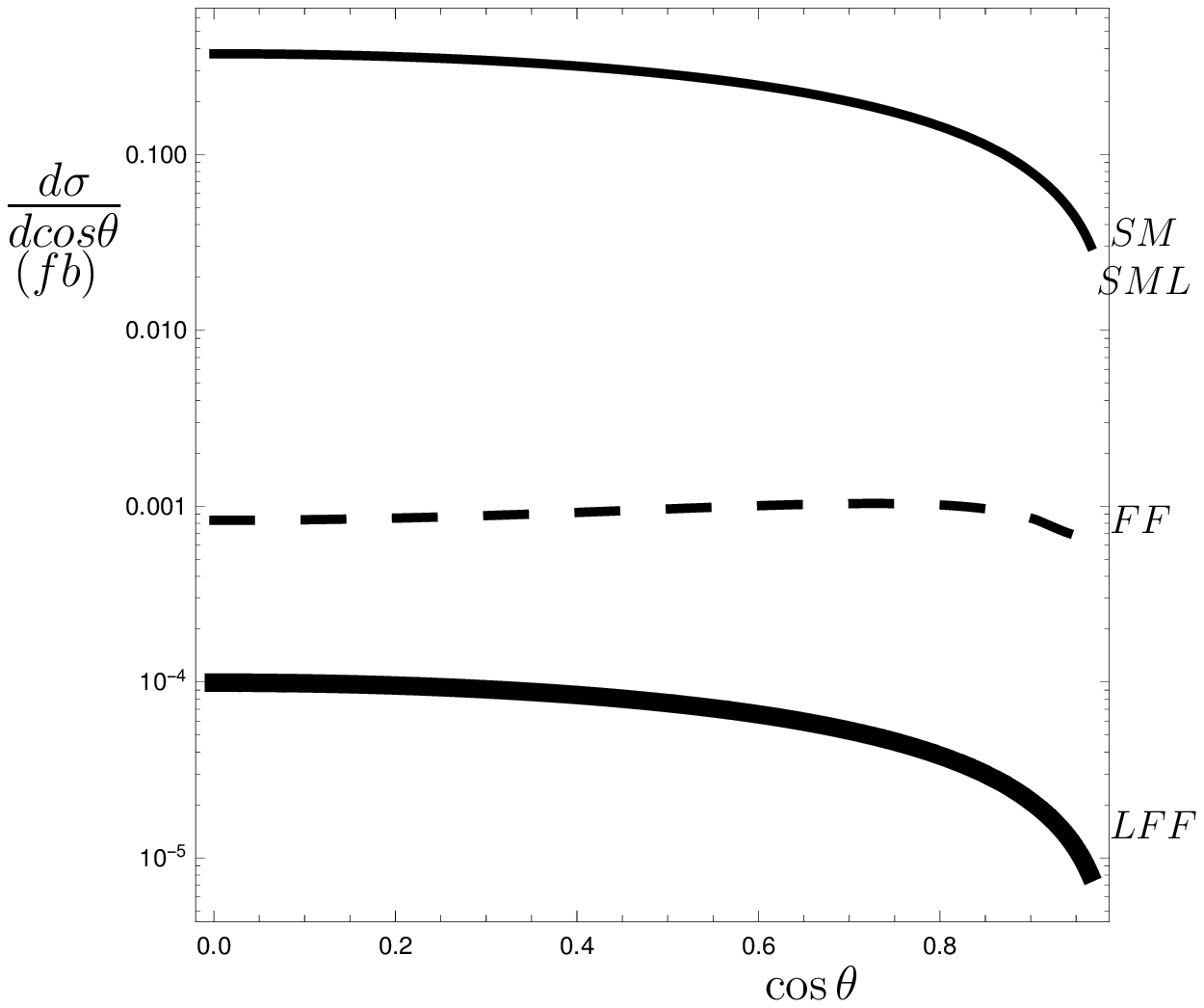, height=8.cm}
\]\\
\vspace{-1cm}
\caption[1] {Energy dependence (upper panel for $\theta=\pi/2$) and angular distribution
(lower panel for $\sqrt{s}=$ 4 TeV)
of the $e^+e^-\to ZH$ cross section. SM refers to the standard
unpolarized case, SML to the standard longitudinal $Z$ production, FF
and LFF to the corresponding cases including the form factor effect.}
\end{figure}

\clearpage

\begin{figure}[p]
\[\hspace*{-1.8cm}
\epsfig{file=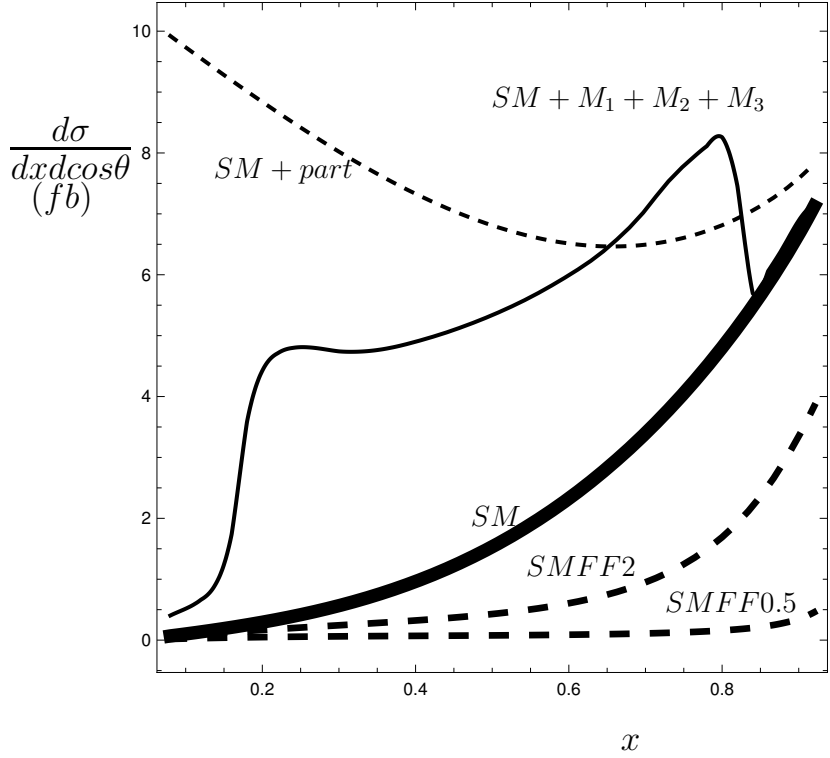, height=10.cm}
\]\\
\vspace{-1cm}
\caption[4] {Inclusive distributions for $e^+e^- \to
Z_L + anything$ for $\sqrt{s}=$ 4 TeV. Comparison of the pure SM contribution 
to the one with form factor effects (SMFF),
to the addition of 3 pairs of new particles and to the addition of a parton-like contribution.}
\end{figure}

\clearpage

\begin{figure}[p]
\[\hspace*{-1.8cm}
\epsfig{file=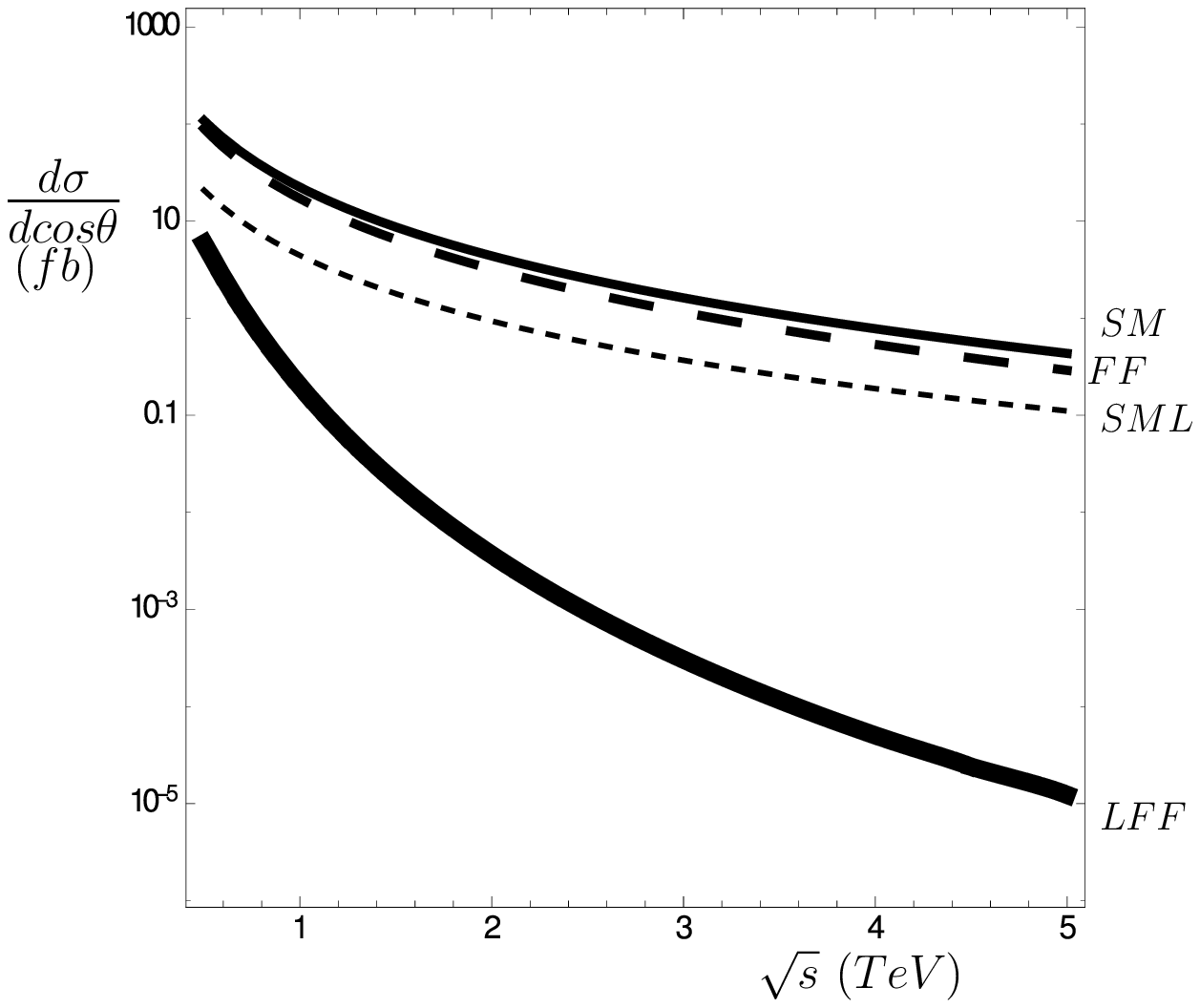, height=6.cm}
\epsfig{file=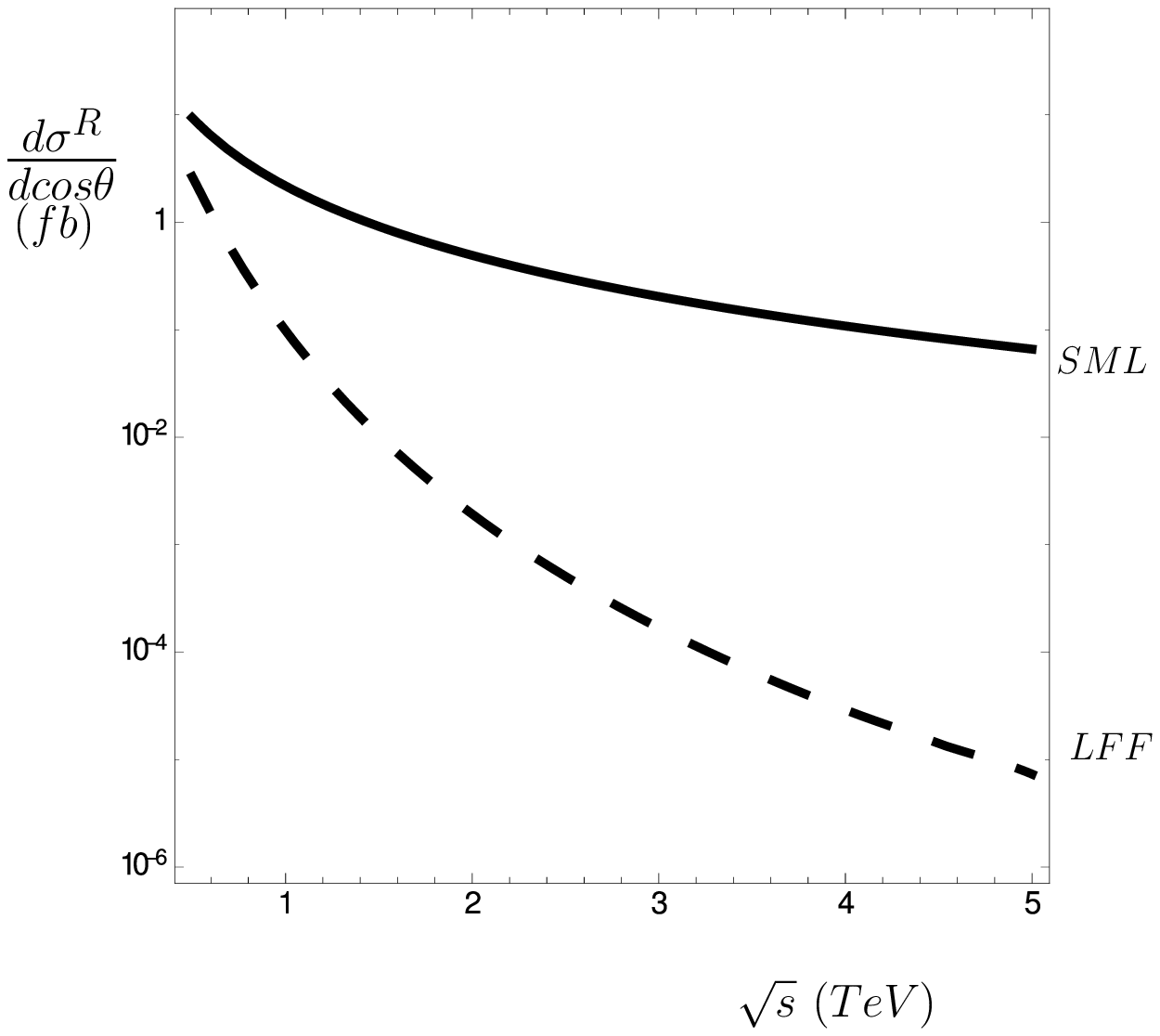, height=6.cm}\]\\
\[\hspace*{-1.8cm}
\epsfig{file=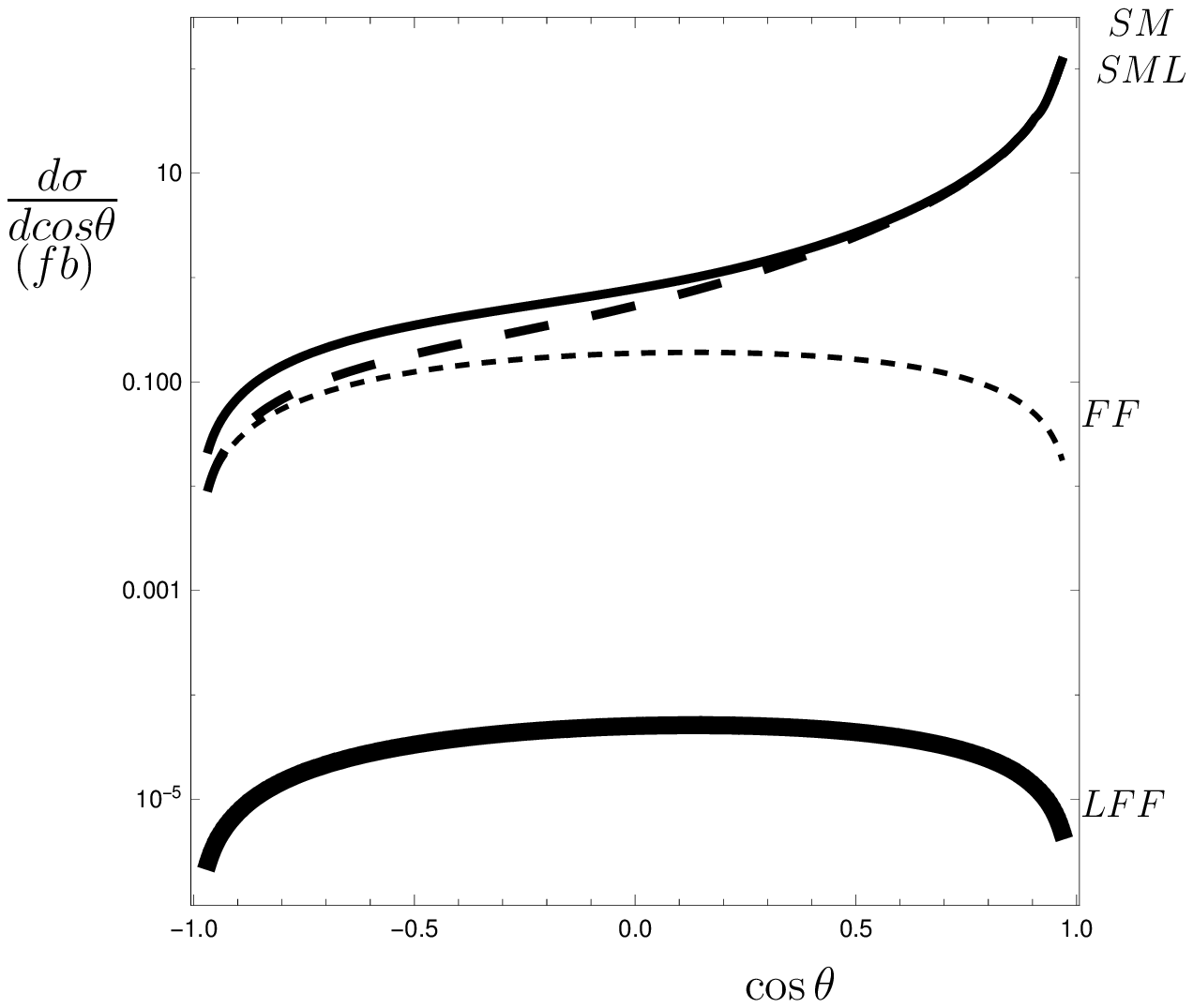, height=6.cm}
\epsfig{file=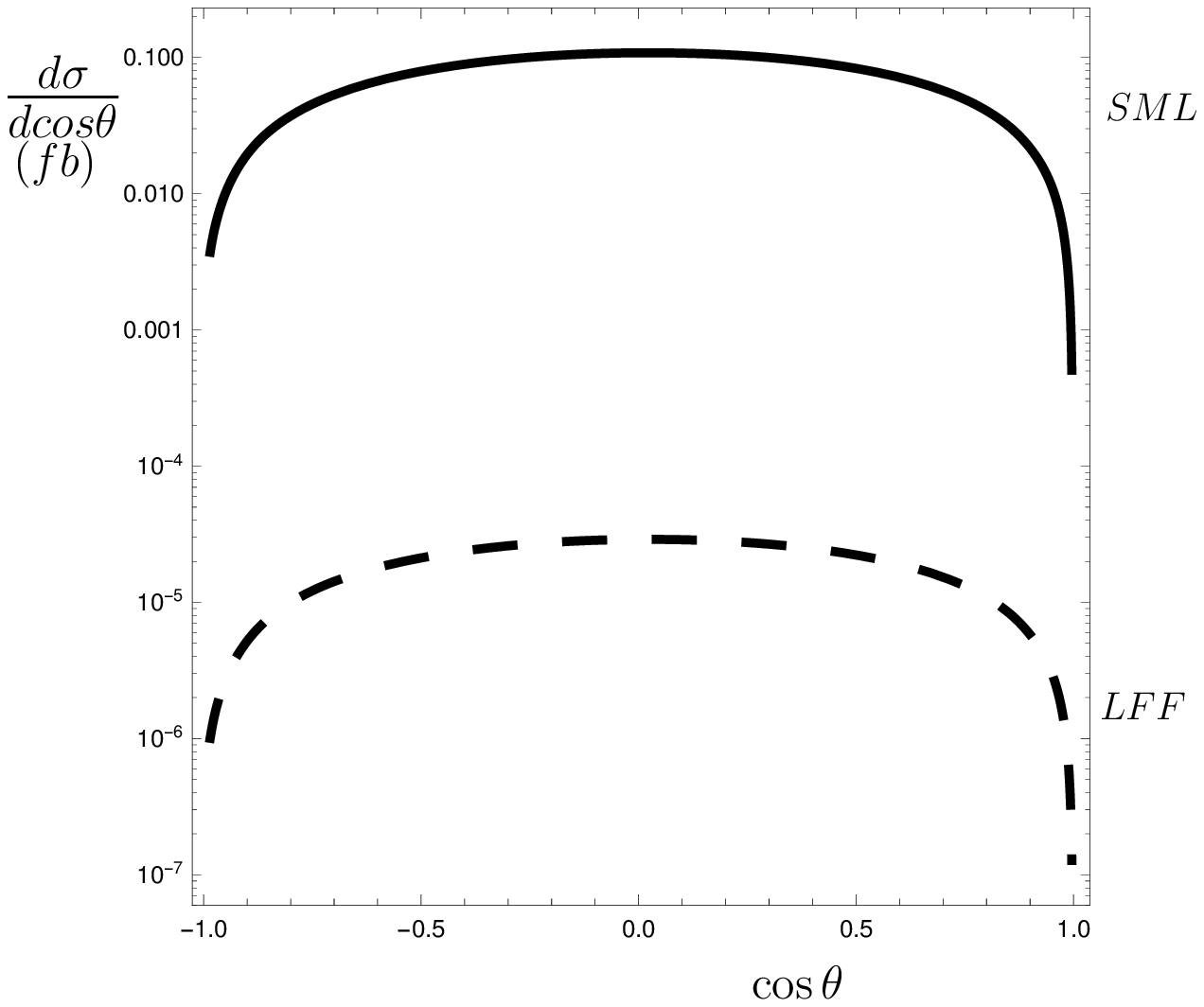, height=6.cm}\]
\vspace{-1cm}
\caption[2] {Upper panel: Energy distribution (for $\theta=\pi/2$) of the
$e^+e^-\to W^+W^-$ cross sections (unpolarized (a) and right-handed polarized $e^-$ (b)) . Lower panel: Angular distribution at 4 TeV of the $e^+e^-\to W^+W^-$ cross sections (unpolarized (c) and right-handed polarized $e^-$ (d)) . Same notations as in Fig. 1 with L now refering to $W^+_LW^-_L$.}
\end{figure}

\clearpage

\begin{figure}[p]
\[\hspace*{-1.8cm}
\epsfig{file=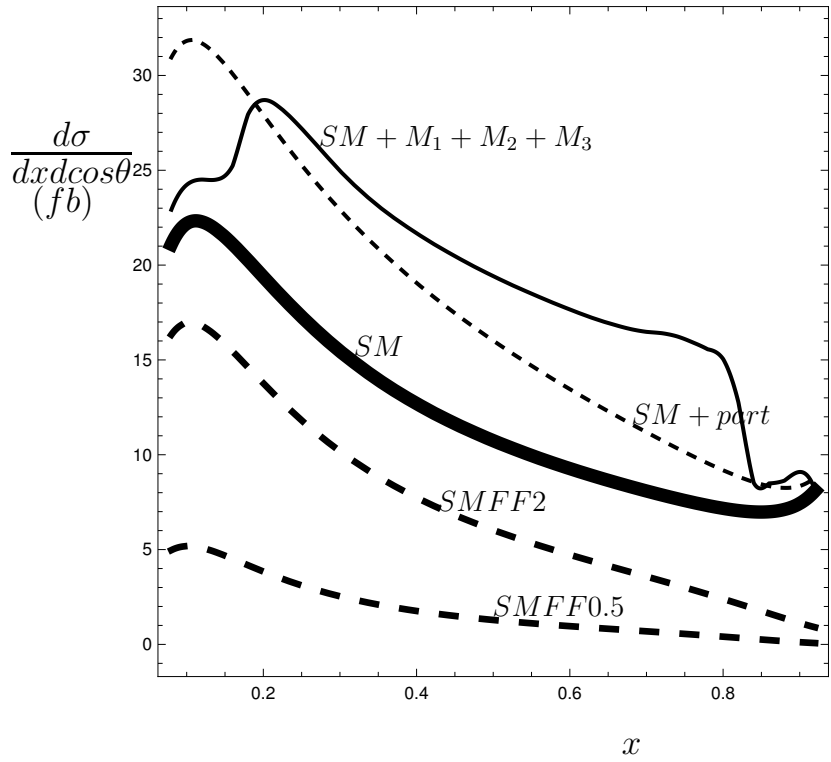, height=10.cm}
\]\\
\vspace{-1cm}
\caption[4] {Inclusive distributions for $e^+e^- \to
W^-_L + anything$. Same comparisons as in Fig. 2.}
\end{figure}

\clearpage

\begin{figure}[p]
\[\hspace*{-1.8cm}
\epsfig{file=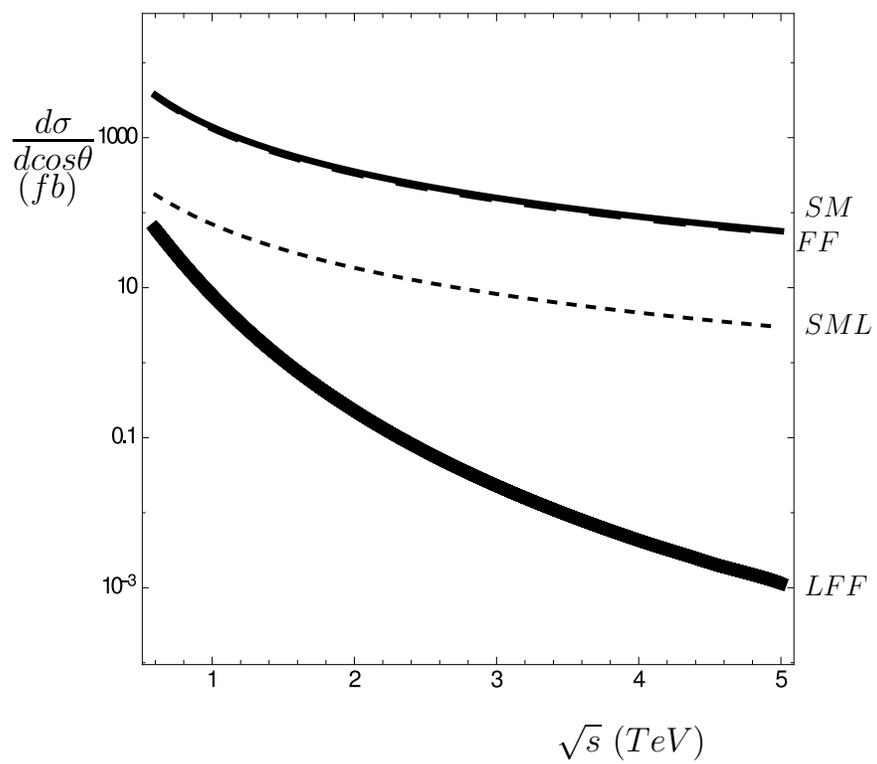, height=10.cm}
\]\\
\vspace{-1cm}
\caption[3] {Energy distribution (for $\theta=\pi/2$) of the $\gamma\gamma\to W^+W^-$ cross section.
Same notations as in Fig. 1 and 3.}
\end{figure}

\end{document}